# Energy Production in Martian Environment
## Powering a Mars Direct based Habitat

## Gianmario Broccia[ab]*

[a]University of Cagliari, Via Marengo 2, Cagliari 09123, Italy
[b]Waseda University, 1 Chome-104 Totsukamachi, Shinjuku City, Tokyo 169-8050, Japan

ABSTRACT

This study investigates the possibility of producing energy on Martian soil assessing what the best energy system would be under given constraints. This goal is contextualized with the adoption of the well-known Mars Direct architecture and the purpose to power a long stay surface habitat with a crew of four astronauts over a Martian year.

The study was carried out by assessing the thermal loss of the habitat, the power consumption of a basic ECLS system, and the producible power according to the environmental conditions.

Both a full-nuclear and non-nuclear configuration of the energy system have been compared, obtaining a clear result.

## 1. Introduction

Mars has always been an object of interest to the scientific community and several mission concepts have existed even before man flew out of the Earth's atmosphere. Among all the concepts that have seen the light over the years, this study refers to the concept of Mars Direct, conceived in 1990 by Robert Zubrin (1), at the time engineer at Martin Marietta.

Mars Direct includes two phases: the landing of the ERV (Earth Return Vehicle) and the arrival of the Lander-Habitat (henceforth "Hab") with a crew of 4 astronauts, two years later. The ERV would spend two years on the surface to produce the fuel for re-entry using the carbon dioxide from the thin Martian atmosphere.

This paper will show how the optimal energy system for the habitat can be chosen. The driving constraint will be the minimum possible mass, assuming to power the base for an entire Martian year under the worst-case scenario, environmentally speaking. To reach the goal, a series of intermediate sub-objectives have been defined:

- Obtaining a heat loss profile for the Hab during the Martian year;

- Simulating an environmental control and life support system;

- Obtaining an electrical and thermal consumption profile for the Hab;

- Assessing which energy sources apply to this context, their producibility, and the best configuration according to the previously introduced constraints;

* Gianmario Broccia. Tel.: +393484665186
E-mail address: gianmario.broccia@gmail.com

The main driver of this study was Nasa's Baseline Values and Assumptions Document (2). This report has been considered for almost all the phases of the work since it provides values and recommendations for the ECLS system, the power system, and so on. All the major considerations of the author have been made within the tolerance proposed by the BVAD.

This study originates from a Master's Thesis and not all the material could have been reported here. Readers are encouraged to contact the Author for any further information.

## 2. Site choice and environmental conditions

To contextualize this study in a precise way it has been necessary the choice of a landing point. As imaginable, the landing site determines an implicit choice of the environmental conditions to cope with. The site has been chosen to maintain a maximum limit of 45° latitude to the north also taking into account that the reference database (Mars Climate Database (3)) relies on data measured by landers and rovers. This consideration led to the choice of the point located at 45° N and 140° E, in Utopia Planitia and very close to the site of Nasa's Viking 2 lander (4). This will ensure reliable data from the database.

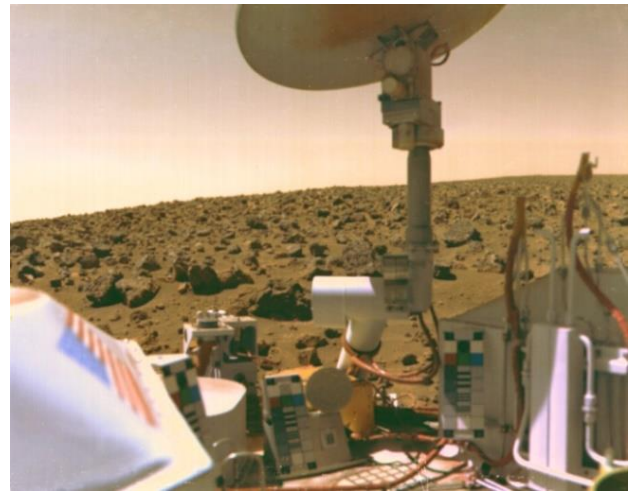

**Figure 1 - Utopia Planitia (1976). Credits: NASA/JPL**

The environmental conditions of this site have been obtained, as anticipated, by the Mars Climate Database (MCD). It was, therefore, possible to have a full characterization of the site with any parameter of interest:

- solar radiation at the ground;

- air and soil temperature;

- atmospheric pressure;

- air density;

- hours of insolation;

- wind intensity.

Furthermore, two main scenarios were considered:

1. **Sun Season**. Average solar radiation, clear sky, and weak winds;

2. **Storm Season**. Sandstorms, almost no solar radiation, and greater wind intensity.

Such storms can last up to two Martian months (5) and yet it has been decided to consider the worst-case scenario in which a storm lasts 6 Martian months, making solar production insufficient. This choice was also forced by the conditions of the MCD, which allowed the "Storm Conditions" only for an Ls (Solar Longitude) between 180° and 360°.

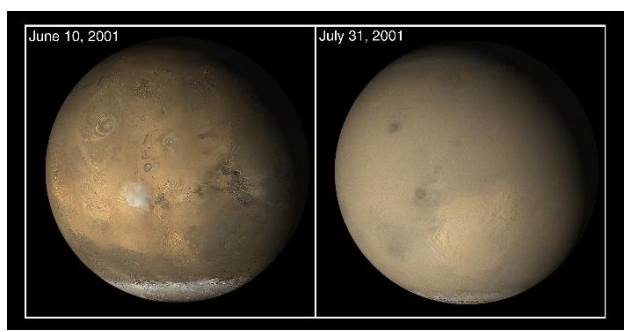

**Figure 2 - Mars Odyssey captures a global dust storm during its arrival. Credits: NASA/JPL-Caltech**

The following table from the MCD shows an example of how some values of interest vary during the two scenarios.

| Value | Air temp. [K] | Wind speed [m/s] | Solar rad. [W/m²] | Air density [kg/m³] |
|-------|-----------|------------|-----------|---------------|
| Sun | 205,7 | 3,5 | 227 | 0,02 |
| Storm | 175,9 | 9 | 11,2 | 0,027 |

**Table 1 – Seasonal averaged values of interest**

The MCD does not allow automatic download of large quantities of data and, considering that the Martian year consists of 669 sols, it has been decided to adopt a characteristic sol for each month that holds the average values for that particular month. For example, the air temperature at 10.00 a.m. in the characteristic sol of the first month is the average value of the temperature in all the sols of that month for that hour.

## 3. Habitat design and Thermal Analysis

The first sub-objective required to deduce internal and external geometry of the Hab, characterization of the walls, the dome, the external armor, and the internal components.

Geometries were derived from Zubrin's publications (6), while for the thermal features it was decided to rely on Nasa's TransHab (7).

With a careful choice of materials, it was possible to recreate a stratigraphy of each structural component of the Hab.

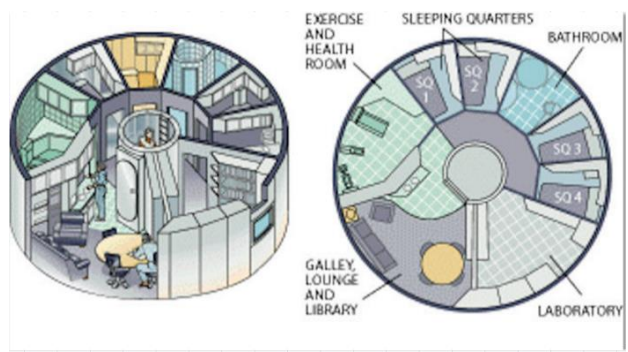

**Figure 3 – First Floor of the Hab and living quarters**

The Hab is a 4,9 m high cylinder with a total diameter of 8,4 m. The outer vertical shell is 21,4 cm thick and the entire structure is supported by landing legs, leaving a space of 1,5 m between the base and the ground.

The thermal analysis was then conducted according to the provisions contained in the Italian Regulation UNI/TS 11300-1:2014 for civil buildings (as a national application of the European regulation EN ISO 13790:2008). The calculation is based on the following thermal balance:

$$Q_H = (Q_{tr} + Q_{ve}) - \eta_H(Q_{int} + Q_{sol})$$

where:

- $Q_H$ is the total net thermal dispersion, through the shell;

- $Q_{tr}$ is the gross loss for thermal transmission;

- $Q_{ve}$ is the loss due to the heating of the ventilation air, here zero since we do not heat the external air;

- $Q_{int}$ and $Q_{sol}$ are internal and solar radiation contributions;

- $\eta_H$ is a coefficient considering the internal thermal capacity of the hab.

The internal contributions, due to machinery and astronauts, are calculated in the part concerning the life support and used to correct the calculations related to the thermal analysis.

It must be specified that the $Q_{tr}$ term also includes two components accounted as "extra flux" by infrared radiation to the sky and the ground and calculated with the Stefan-Boltzmann law applied to the Hab's vertical wall. In the case of the soil, it has been performed a calculation which took into account elements such as optical depth and the view factors between the ground and the vertical wall of the Hab (hereinafter simply named "shell").

The extra flux was corrected with the proper view factor (8), in this case, related to a cylindric structure 1,5 m above the ground (due to the landing legs, indeed).

A separate path was made for the liminar coefficient (still included in $Q_{tr}$), necessary for the calculation of the transmittances for external components (shell, windows, dome). This coefficient, referred to as $h_{lim}$, is the sum of convective ($h_{conv}$) and radiative ($h_r$) coefficients referred to as the shell liminar layer.

To obtain a possible value of the $h_{conv}$, a thermal analysis based on a model of the Phoenix lander was considered (9). Comparing the environmental conditions between Phoenix's site and the one being studied here, and considering that the $h_{conv}$ depends on the Nusselt number (10) which shows a negligible variation between the sites (0,45% on average), Phoenix's value can be adopted in this case as well.

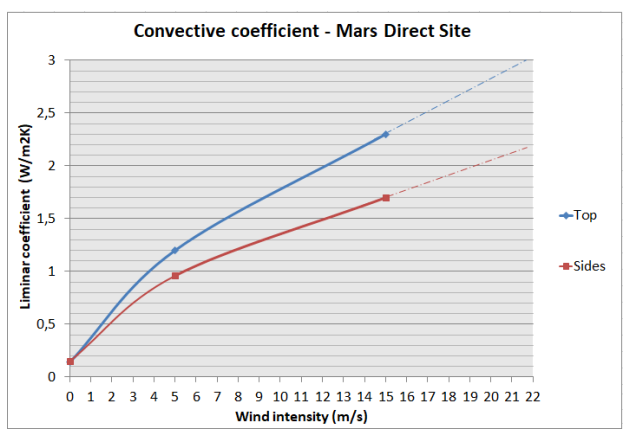

**Figure 4 - Convective Coefficient from Phoenix's site for the horizontal and vertical shell of the Hab respect windspeed**

The radiative coefficient $h_r$, on the other hand, was calculated using the shell temperature, according to the following relation (UNI EN ISO 6946-2007):

$$h_r = 4\varepsilon\sigma\left(\frac{T_{shell} + T_{air}}{2}\right)^3$$

where the quantity in brackets is called "film temperature". It has been assumed that the insulation of the shell is the most effective possible, which means that the shell temperature approximates $T_{air}$.

With this consideration, it is possible to have an average of the liminar coefficient to apply to the stratigraphies.

The global transmission coefficient $H_{tr}$ (W/K) is strictly dependent on the liminar coefficient which knowledge allows the calculation of the $Q_{tr}$ in function of the temperature gradient, according to the UNI/TS 11300 Standard:

$$Q_{tr} = H_{tr}(T_{int} - T_{air})$$

Now the monthly average profile that shows the variation of losses along the year and the free contribution (in blue) can be obtained.

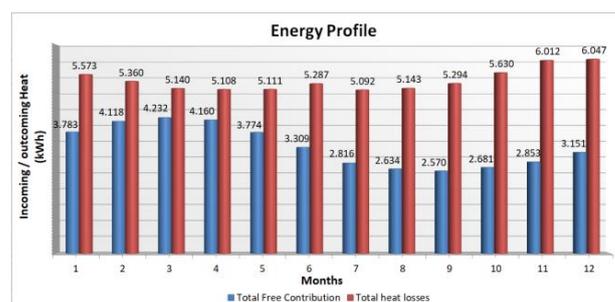

**Figure 5 - Monthly Average heat loss and free contributions**

Another representation has been made to show the loss over each characteristic martian sol.

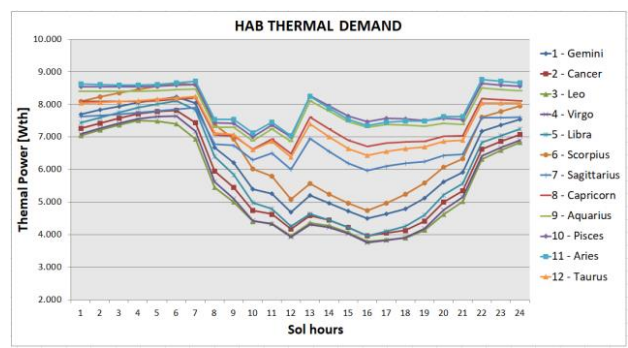

**Figure 6 - Daily heat loss on 24 hours for each month**

A difference in the noon hours can be seen between the months in sun season and storm season. The second ones show a higher heat loss, in fact.

The warmer months are on the left side of the graph because the data are data acquired from a solar longitude equal to 0, a value that corresponds to a period set towards the end of the Martian spring.

## 4. Environmental Control and Life Support

### 4.1 Introduction

This part has been largely based on a previous study dated 1993 and realized by Chatterjee Sharmista (11). This study has been conducted with the Aspen program, as well as this one, and provided fundamental indications for the realization of a model suitable for this study. Given the age of Sharmista's study, some points have been updated, generating quite divergent final values.

### 4.2 Crew Metabolic Activity

The first goal was to determine the metabolism of the crew, therefore the need for oxygen, food, water, as well as the emission of carbon dioxide, and other wastes. To do this, unlike Sharmista, it was decided to start from tables that report known values of latent and sensible heat emitted by the crew (12). It was also necessary to estimate a decrease in metabolic rate (MR) due to reduced gravity. Knowing the MR

on a Shuttle mission (13), the corresponding value in Martian gravity has been calculated this way:

$$MR_{mars} = MR_{earth}(0,57 + 0,43K)$$

where K is the gravity acceleration of the considered celestial body respect Earth. In the case of Mars (38% of Earth's gravity), the metabolic rate would be, in first approximation, equal to 73,3% of normal. The values related to the sensitive and latent heat emitted by the crew have therefore been corrected according to this parameter.

| Activity | hours/sol |
|----------|-----------|
| Sleep | 8 |
| Nominal activity | 12 |
| Heavy exercize | 1 |
| Rest | 3 |

**Table 2 - Hours of activity**

Finally, a division of the hours of activity was decided to calculate the daily energy requirement and the total sensible heat emission, to set the $Q_{int}$ value in the thermal analysis. Regarding this last point, it has been assumed that a single woman/man couple trains during an hour of exercise.

| Parameters | Males | Females |
|------------|-------|---------|
| Body mass (kg) | 75 | 60 |
| Height (m) | 1,83 | 1,63 |
| **Age (y)** | 35 | 32 |
| **MR Broccia (kJ/CM*sol)** | 10.806 | 9.302 |
| **MR Sharmista (kJ/CM*sol)** | 12.958 | 11.729 |

**Table 3 – Daily Metabolic Intake**

It is possible to appreciate a comparison between both studies in Table 3. The difference consists in having considered reduced gravity and much more updated tables than the work considered by Sharmista (14).

Respect Sharmista, an increase in the intake of 25% has been considered assuming it is dedicated to motor activities only and does not involve the emission of sensible or latent heat. The daily energy requirement was used to calculate the food needed to sustain the crew. These substances and their metabolic reactions have been considered likewise Sharmista as:

- Glucose, in place of carbohydrates

  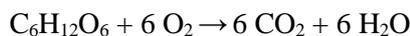
  $$C_6H_{12}O_6 + 6\,O_2 \rightarrow 6\,CO_2 + 6\,H_2O$$

- Alanine, in place of proteins

  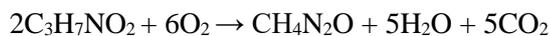
  $$2C_3H_7NO_2 + 6O_2 \rightarrow CH_4N_2O + 5H_2O + 5CO_2$$

- Palmitic Acid, in place of fats

  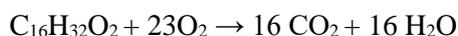
  $$C_{16}H_{32}O_2 + 23O_2 \rightarrow 16\,CO_2 + 16\,H_2O$$

- Cellulose, in place of fiber, does not take part in any reaction.

Considering that Sharmista includes a partial absorption of the body (97%, 95%, 92%, 0% respectively for the nutrients listed above) and therefore a surplus of incoming food is needed, it has been possible to calculate the oxygen and the emission of carbon dioxide from simple stoichiometric calculations, considering reactions as the only processes involved. Similarly, considering the production of Urea, a corresponding value of urine was calculated.

The proportion of unabsorbed food is simply considered to be fecal matter.

| Total Inputs and Outputs (kg/sol) | |
|---|---|
| **Food need** | 2,27 |
| **Oxygen** | 2,83 |
| **Drinkable Water** | 11,38 |
| **Metabolic Water** | 1,43 |
| **Rehydr. Food water** | 3 |
| **Water consumption** | 14,39 |
| **Carbon Dioxide** | 3,34 |
| **Persp. Water** | 5,41 |
| **Urine** | 10,04 |
| **Solid Feces** | 0,25 |
| **Fecal Water** | 0,37 |

**Table 4 - Crew Inputs and Outputs**

### 4.3 Cabin Ventilation

The ventilation flow is a fundamental point to guarantee constant conditions in the cabin (24°C and 50% humidity). Considering the presence of contaminant inside this closed-loop cycle, the flow rate must also be modulated according to the molar fractions of its components so that the composition of the air entering the cabin remains ultimately equal to:

- 0,00033 Mol-frac $CO_2$
- 0,77495 Mol-frac $O_2$
- 0,21000 Mol-frac $N_2$
- 0,01472 Mol-frac VAP

The flow rate has been calculated according to typical reasoning and relationships of psychrometry, setting a constant intake temperature equal to 40 °C and wanting a flow rate that balances the sensible heat loss at any time, also taking into account the increase in specific humidity and the sensible heat that the crew brings into the cabin.

Using a respiratory model that included a simulation performed in Aspen Plus, it was possible to derive the flow and the composition of the air exhaled by the astronauts, having known the flow rate and the composition equal to that in the cabin.

It has been possible to derive the characteristics of the incoming air using a simple mass balance, considering the cabin as a control volume.

It has also been considered the presence of a hot battery which purpose is to bring the air to the intake temperature of 40 °C.

| Airflow to cabin | | | |
|---|---|---|---|
| **Wt; kg/h** | **Min** | **Mean** | **Max** |
| **Mass Flow** | 494,79 | 902,82 | 1175,32 |
| **Mol-frac $CO_2$** | 0,000147 | 0,000253 | 0,000224 |
| **Mol-frac $O_2$** | 0,775094 | 0,776848 | 0,775355 |
| **Mol-frac $N_2$** | 0,210129 | 0,210726 | 0,210233 |
| **Mol-frac VAP** | 0,012277 | 0,014525 | 0,014188 |
| **Hot Battery** | 3798,1 | 6816,6 | 8810,6 |

**Table 5 – Some parameters related to the airflow in cabin**

### 4.4 Subsystems

Life Support consists of several subsystems briefly explained below. A full explanation can be found, again, inside Sharmista's work.

**Vapor Compression Distillation (VDC)**

It consists of a system that aims to recycle water from urine, hygiene, and feces. This sorting system is a distiller that eliminates the waste components and performs separation when the water is in the vapor phase. Subsequently, pure water is compressed to the liquid state, while the latent heat is recycled and sent to the evaporator through a multistream exchanger (which works both as a condenser and as an evaporator). Water is returned to atmospheric pressure with a pump.

In this work, this system processes about 39,8 kg/sol of water and does not require a particular power intake, except that coming from the compression and which is slightly higher than 1 kWe.

**Air Revitalization System (ARS)**

This system acts to eliminate the contaminants in the cabin. Such contaminants are shown below.

| Contaminants Production (g/sol) | |
|---|---|
| CO (Carb. Monoxide) | 0,5125000 |
| CH$_4$ (Methane) | 0,0512500 |
| C$_2$H$_6$ (Ethane) | 0,0003417 |
| C$_2$H$_4$ (Ethylene) | 0,0000085 |
| C$_2$H$_2$ (Acetylene) | 0,0000854 |
| H$_2$ (Hydrogen) | 0,0444167 |
| Total Flow | 0,6086 |

**Table 6 – Contaminants generation in cabin**

The air coming from the cabin is separated into a special separator with an absorbent bed. Here the contaminants are separated from the air and sent to appropriate reactors where they are oxidized through a quantity of oxygen (therefore air) stoichiometric. The reaction is considered 100% efficient and the carbon monoxide is oxidized in the reactor at room temperature, producing carbon dioxide, while the remaining contaminants are oxidized in the high-temperature reactor (450 °C), producing carbon dioxide and water vapor. The air and the reaction products are then mixed and sent to the next subsystem.

**Humidity Control**

In this phase, there is the elimination of the water introduced into the cabin by breathing and perspiration, as well as the water generated in the previous subsystem.

The subsystem counters the variation in specific humidity and therefore returns to the cabin value (about 9,34 g/kg).

The system is modeled with a heat exchanger where the air is saturated at an average temperature of 12,6 °C, while the excess heat is disposed towards the external environment, being low heat temperature. A subsequent separator, which also acts as a condenser, eliminates the excess water from the airflow. A process of this type is generally carried out using surfaces at a lower temperature on which the water is condensed.

The amount of water separated in an hour depends on the activity that the crew is carrying out. The maximum value, for example, is typical of the two consecutive hours in which the crew performs the physical activity.

**Humidity Control Parameters**

| Wt; g/h | Min | Mean | Max |
|---|---|---|---|
| Expelled Heat | 5371,6 | 2379,1 | 3803,7 |
| Removed Water | 141,53 | 266,2 | 770,25 |
| Coolant Flow | 247,89 | 396,32 | 559,68 |

Table 7 - Parameters of the humidity control

## CO₂ Removal and Reduction and O₂ Production

In this subsystem, carbon dioxide is separated from the air and the production of oxygen takes place. The airflow is first sent to a separator that represents an absorbent bed and then sent to the actual subsystem. This apparatus includes two main blocks: the Sabatier reactor and the electrolyzer.

The Sabatier reactor uses the homonymous reaction to produce methane and water from carbon dioxide:

$$CO_2 + 4\,H_2 \rightarrow CH_4 + 2\,H_2O$$

The hydrogen can be produced by electrolysis from wastewater that the ERV produces during the synthesization of fuel (methane/oxygen) with the same reaction. Regardless of its use, methane is considered expelled from the system "Hab".

The water is sent to the electrolyzer to obtain the oxygen to be re-injected into the air and hydrogen to be recycled to the Sabatier reactor:

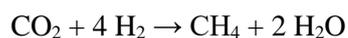

$$2\,H_2O \rightarrow 2\,H_2 + O_2$$

Observing the proportions of the reactions it can be seen that the hydrogen that is recycled to the reactor amounts to half of the total flow entering the reactor, in the hypothesis of reactions with ideal efficiency. The rest will, therefore, be an external feed. A part related to the CO₂ surplus from the atmosphere is also taken into account.

**Reactor and Electrolyzer flows (g/h)**

| | |
|---|---|
| CO₂ treated | 164,3 |
| H₂ feed | 30,1 |
| CH₄ produced | 51,05 |
| Produced Water | 132,19 |
| O₂ produced | 119,46 |

Table 8 – Flows for the CO₂ reduction and O₂ production subsystem

On the other hand, the oxygen is sent to a mixer and put back into the air. It must be noted that the oxygen produced during the whole process is not sufficient to cover the needs of the crew. This implies that about 24,23 g/h of CO₂ must necessarily be taken from the external environment.

For this reason, a simulation with a generic compressor with exchanger has been carried out, aiming to compress the external CO₂ from the average atmospheric pressure of 0,0085 bar up to the reactor pressure (1,013 bar), while the exchanger has the task to cool the gas to the required temperature (300 °C). The values of power related to the compressor and to the transferred heat by the exchanger to the environment are however negligible (even if they have been equally considered in the final requirement).

The Sabatier reactor has been considered with a total conversion of the reactants, a fact that could be considered not far from the reality since NASA showed that conversion of 99,9% was obtained (15). The temperature was instead chosen according to an ENEA study (16) that shows how the greatest efficiencies can be achieved by making the reactor work at about 300 °C.

For electrolysis, a membrane electrolyzer (PEM - proton exchange membrane) with an operating temperature of 80 °C at atmospheric pressure has been considered. Not being able to perform an in-depth simulation due to lack of time, data and not being able to model the electrolysis, it was decided to use a simple reactor together with a separator and to calculate the power required by Excel. It was decided to consider a "classic" value of 55 kWhe/kg of hydrogen produced. The absorbed power (constant) is equal to about 1,63 kWe.

## 5. Thermal and Electric Demand

The total requirement was calculated considering several actors: circulation pumps, heat exchangers, compressors, fans for air circulation (17), hot battery for cabin air, lighting (18) and equipment taken from a Caltech analysis on its laboratories (19).

Having no data available, it has been decided to consider the thermal emission by the equipment (including PC, laundry and scientific equipment) numerically equal to 60% of the electricity needs.

From the results, it can be seen that this type of equipment has peaks that reach 5 kWe, while during the night the consumption remains constant and equal to just under 2 kWe.

Taking into account all the elements described above, we obtain a final balance (Fig.7) which can be summarized with the following values.

| Consumption | | | |
|---|---|---|---|
| **We; Wt** | **Min** | **Mean** | **Max** |
| **Elec. Consump.** | 6118,3 | 9294,6 | 12264,6 |
| **Ther. Consump.** | 3760,5 | 6778,9 | 8772,3 |

**Table 9 – Final consumption**

It, therefore, appears that the electrical requirement is characterized roughly by the same profile throughout the year, except for the values, which show variations up to 1 kWe, especially in the time slot between 15.00 and 20.00 (martian hours).

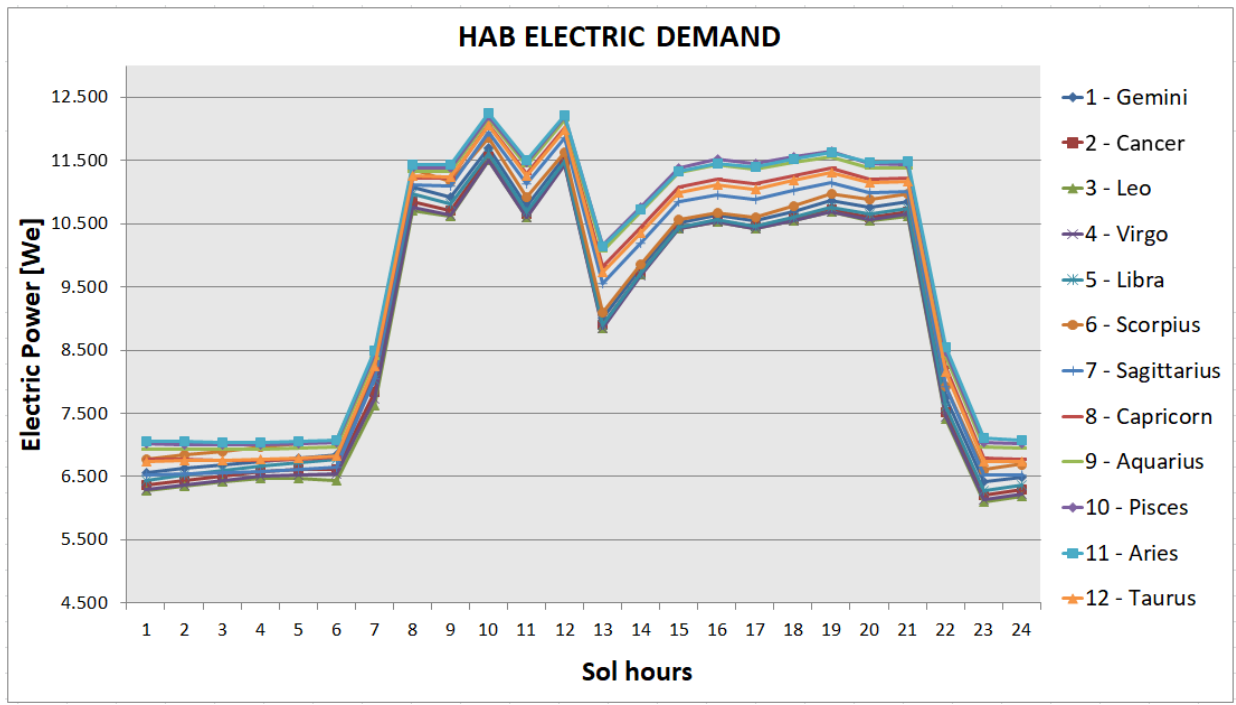

**Figure 7 - Electric demand along the year**

The maximum value is about 12,3 kWe while the night consumption remains on an average of 7 kWe.

## 6. Energy Sources and Producibility

Before introducing the possibilities for the generation in the Martian environment, it is useful to recall the typical conditions of Mars:

- Atmospheric pressure is almost 1% of the Earth's.

- The regolith can cover the solar panels and erode unprotected mechanical devices.

- The solar radiation is about 2-3 times lower than on the earth, but the thin atmosphere allows to take full advantage of the radiation.

- Sandstorms can block the majority of solar radiation, making photovoltaics practically useless. However, this leaves a chance for wind turbines.

### Eolic

Wind power is still a controversial source. The studies that consider it appear to be in a number equivalent to those that do not and, in general, it seems that a turbine specifically designed for Mars has not yet been tested. For this reason, this study proposes a comparison between a conventional turbine, the NPS100-24 Artic by Northern Power Systems, and an airborne turbine that was considered in 1999 by Nasa's study (20).

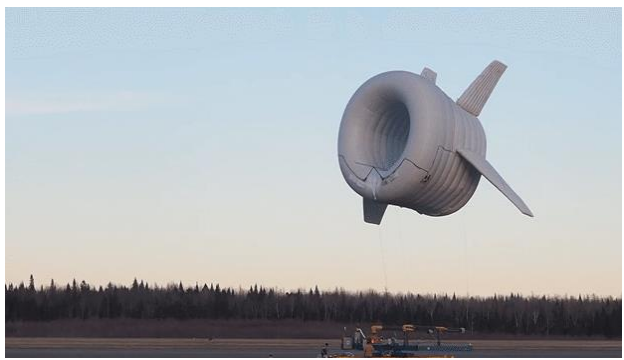

**Figure 8 – Turbine developed by Altaeros Energies**

The power curve of the NPS100-24 Arctic has been supplied by the manufacturer itself (the Northern Power System is no longer operative and the brochure this work relied on can be requested directly to the author), while the power curve of Nasa's turbine is derived from the Betz's law:

$$P_{wt} = \frac{1}{2} C_{pr} A \rho v^3$$

It has been also necessary to take into account that the cut-in speed is higher than on Earth due to the thinner atmosphere. This means that a higher wind will is necessary to generate the same amount of dynamic pressure on the rotor.

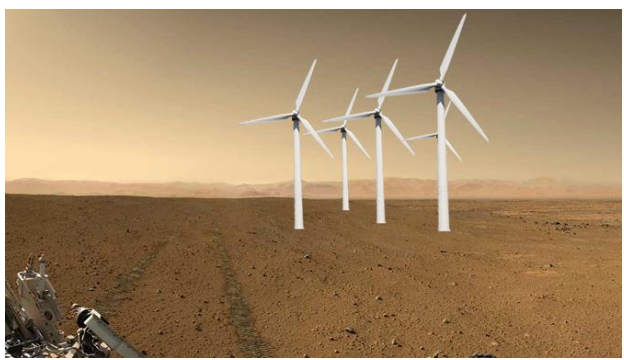

**Figure 9 – Representation of conventional turbines on the surface of Mars**

This condition was considered by comparing the dynamic pressure generated by the Martian wind with the dynamic pressure that occurs in standard terrestrial conditions with a wind of 3 m/s (typical cut-in value for many turbines). With an Excel "IF" function, the command to calculate the power generated by the turbine has been set only for cases in which the Martian dynamic pressure is greater than or equal to the terrestrial pressure at 3 m/s. The dynamic pressure was calculated as follows:

$$q = \frac{1}{2} \rho v^2$$

It is possible to compare the intensity of the Martian wind with Earth's to find an equivalence that makes easier the understanding of the conditions present on Mars. This equivalence is always produced at the same dynamic pressure:

| Comparison between windspeeds (m/s) | |
|---|---|
| **Mars** | **Earth** |
| 1 | 0,1 |
| 2 | 0,3 |
| 4 | 0,6 |
| 6 | 0,8 |
| 8 | 1,1 |
| 10 | 1,4 |
| 12 | 1,7 |
| 14 | 1,9 |
| 16 | 2,2 |
| 18 | 2,5 |
| 20 | 2,8 |
| 22 | 3,1 |
| 24 | 3,3 |
| 26 | 3,6 |
| 28 | 3,9 |
| 30 | 4,2 |

**Table 10 – Equivalent terrestrial windspeed**

It is clear that wind turbine needs very high wind intensity to be useful and this brings the cut-in speed from the classic 2-3 m/s no less than 14-16 m/s, considerably narrowing the range of useful intensity, remembering that on the ground no values above 22 m/s were observed in the MCD.

To calculate the generated energy (to be inputted into the grid), a rotor power factor of 0,4 and an electromechanical efficiency of 98% were considered.

The resulting profile of the NPS100 shows a very poor generation, concentrated in a few hours a sol in relation only to the two months of Aquarius and Pisces, barely reaching a peak of 2,5 kWe out of a nominal output of 100 kWe.

The airborne turbine, on the other hand, manages to generate an electric output that is much better suited for the needs of the Hab.

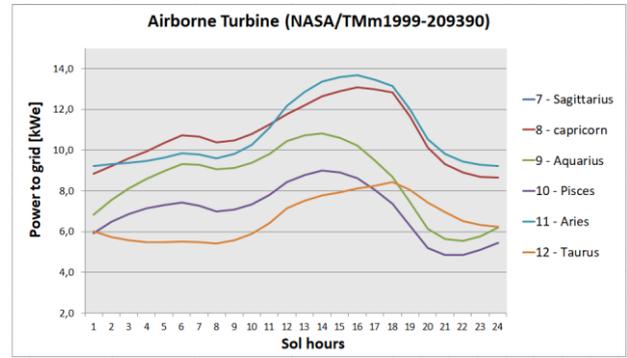

**Figure 10 – Example of profile with a 3m-bladed wind turbine**

This turbine would be suspended at 8000m, benefiting from much stronger winds and lower stratification of the Martian atmosphere. It must be said that the lower gravity leads the atmosphere to have a density variation as a function of the height much less marked than on Earth.

**Photovoltaics**

For photovoltaic purposes, an ideal efficiency of 28%, typical of GaInP/GaInAs/Ge cells, was considered. It has therefore proved necessary to calculate the operating temperature of the cells under Martian conditions, an operation facilitated by the work of Bonal and Torres (21), which provide the following simplified relationship, highlighting its good accuracy in comparison to more complicated models:

$$T_c = 1,00116 T_a + 0,0313174 \Phi_s - 0,108832 u$$

Where $T_a$ is the air temperature, $\Phi_s$ is the radiation flux and u is the wind speed.

In this way, it has been possible to calculate an average operating temperature of about 220K during the hours of light and in the months of the sun season. This parameter was therefore useful for the calculation of the real efficiency under martian conditions. It was also decided to consider a 2% loss due to sand accumulation on the panels and a loss of 5% due to line resistance.

The presence of a solar tracker was not considered due to the lack of data about the mass of similar devices and their actual functionality in an environment with fine particles such as the martian regolith. The plant efficiency was thus found to range between 25,1% and 26,6%.

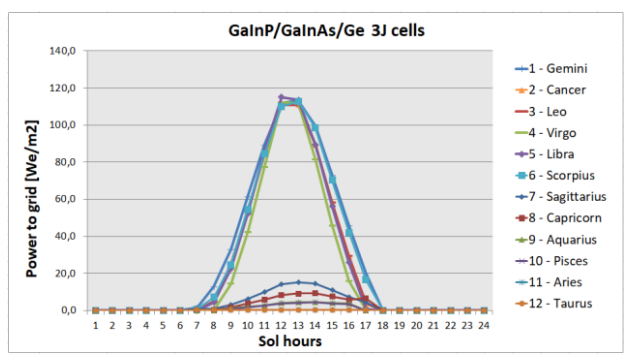

**Figure 11 – PV producibility in sun and storm season**

Fig.11 shows how it is possible to reach peaks of 120 W/m², thanks to the high efficiency of the panels. Another important and immediately visible factor is how the producibility changes in the storm season. The maximum achievable value drops to about 20 W/m² and this makes it impossible to rely entirely on photovoltaics.

**Nuclear Reactor**

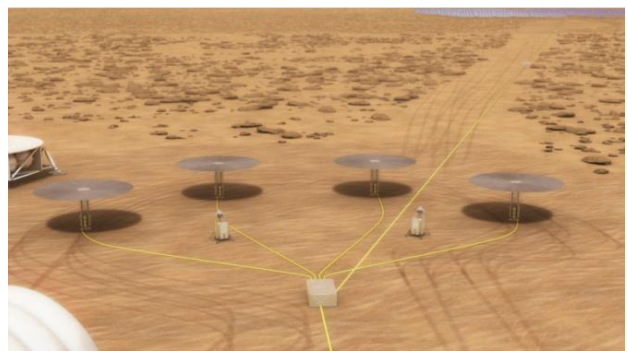

**Figure 12 – An example concept of four Kilopower**

The considered reactor is currently being studied at the NASA Glenn Research Center (22) and it is available in small versions of 1, 3, 5, 7, and 10 kWe with an electrical efficiency of 23%, while the thermal efficiency (for the exploitation of the waste thermal energy) has been considered assuming inevitable thermal losses:

$$\eta_{th} = 0{,}6 \cdot (1 - \eta_e) \approx 0{,}462$$

**RTG**

To meet the thermal requirement, deciding not to use a reactor, it was also considered the GPHS-RTG (23), an RTG now widely tested on different exploration probes such as Galileo, Cassini, and New Horizons. This RTG produces 3500 Wt net using about 7,8 kg of Pu[238] and counting a thermal efficiency of 80%.

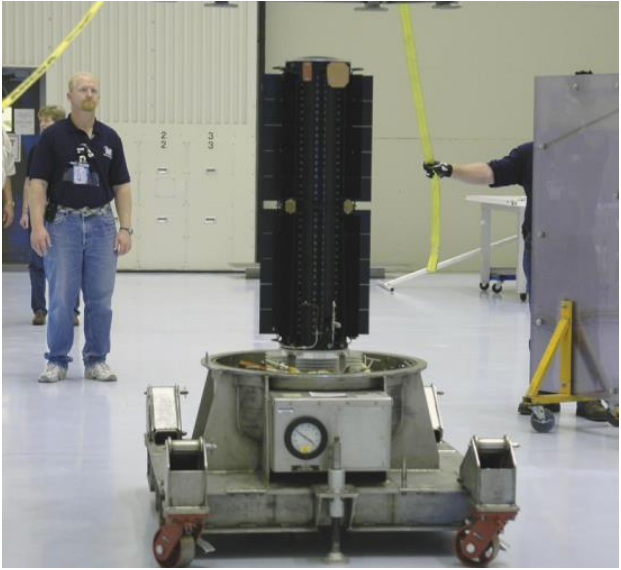

**Figure 13 - General Purpose Heat Source**

### PEM Fuel Cells

PEMs (proton exchange membranes) Fuel Cells use a membrane as an electrolyte and allow protons to pass from the anode to the cathode (24). The protons are obtained from the hydrogen used as a feed, while the electrons contribute to generating current.

Such cells are particularly interesting because they produce simple pure water as waste, which can be used in an electrolyzer in a recycling loop.

The characteristics of the Fuel Cell were assumed by an NREL study (25)  and consider a generation of 20 kWh per kg/h of hydrogen acquired from the storage, assuming that this apparatus can convert the 60% of the energy associated with the lower calorific value of hydrogen (33 kWh/kg).

The BVAD 2015 provides a value of the mass/energy ratio, which is equal to 4 kg/kWhe,

and considers such fuel cells as a good choice with a good technological readiness.

### Batteries

The batteries were considered inspired by two Tesla products: PowerWall2 (26) and PowerPack (27). The first is a purely domestic battery, with 13,5 kWh of capacity and a nominal power of 5 kW for 122 kg, while the second is a battery of more important size with 210 kWh of capacity and a nominal power of 50 kW for 1622 kg. A 100% charge/discharge efficiency has been considered.

### Inverter

That BVAD 2015 does not recommend any particular point for the energy conversion and thus an inverter has been chosen according to the best product from the market. It must be pointed out that no models have been chosen in particular, but rather the best features have been considered.

Considering three commercially available products (Sunny Tripower by SMA, Solivia 11TR by Renelecton, and PVI-12.5-TL-OUTD by ABB) the chosen mass/power factor is thus 3 kg/kWe, while the efficiency is set equal to a classic 99%, typical of all those components. The rated power will be however considered to be enough to couple the generation with the load, even though the value given by the manufacturers is slightly lower than our maximum electric demand (almost 12,25 kWe).

## 7. Energy System Configurations and results

### 7.1 Constraints and Configurations

Once all the available energy sources are known, it is necessary to understand what might be the best configuration, assuming some conditions. The fundamental constraint of the optimization process is the lowest possible mass, while secondary constraints are:

- The constant load balancing, both for electrical and thermal power.

- Storage level always less than or equal to the maximum capacity.

- Initial value of non-zero accumulation (to account that the time horizon begins at night in the sun season when PV and wind power cannot produce).

- Nuclear, Fuel cells, and RTG exert a constant output. The RTG, in particular, can provide thermal energy for a non-nuclear configuration.

- The need to find an alternative power source during the 7$^{th}$ month of Sagittarius (first month of storm season), characterized by the absence of both proper sunlight and enough wind to start the turbine.

Remaining in the domain of simple assumptions, two configurations have been planned: a full-nuclear and non-nuclear.

The driver for the following results was indeed the relation between electric output (kW) and the mass of the related source (kg). Such "specific power" can be considered as a penalty factor. Such factors have been mainly found within the BVAD 2015.

Both configurations brought the necessity of sizing a battery and the main tasks have been two: assessing that the charge/discharge power was within the Powerwall or PowerPack (5kW) capabilities and the needed capacity to follow the load.

### 7.2 Full-Nuclear Configuration

In this case, the configuration is much simpler and related to the entire Martian year. The reactor is coupled with a battery on the model of the PowerWall 2, which is used to follow the variations of the load, while the reactor provides a constant power of 10 kWe. This choice allowed a to have a 1575kg reactor instead of a 10+3 kWe electric configuration (necessary to cover the top electric demand without a battery) that would have had a mass of 2296 kg with the inconvenience of producing much more energy than necessary over the entire year.

This configuration makes it possible to satisfy both thermal and electrical loads thanks to a Stirling engine that exploits the waste heat (28) (29). Despite the cited papers show a higher electric efficiency for the engine, the author preferred to remain on a more "commercial" value of 23%, to work in the worst-case scenario.

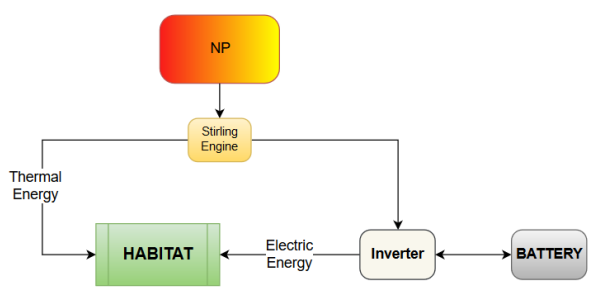

**Figure 14 - Full Nuclear Configuration**

The second issue has been solved with the following relation:

$$Charge(t) = Charge(t-1) + P_{bt}(t)$$

where $P_{bt}(t)$ is the charged/discharged energy during the t hour. The level of charge has been constrained with an upper bound equal (capacity of the battery), set arbitrarily with an IF condition in Excel.

This value has been set equal to the capacity of the PowerWall 2 (13,5 kWh), returning the first raw profile here below.

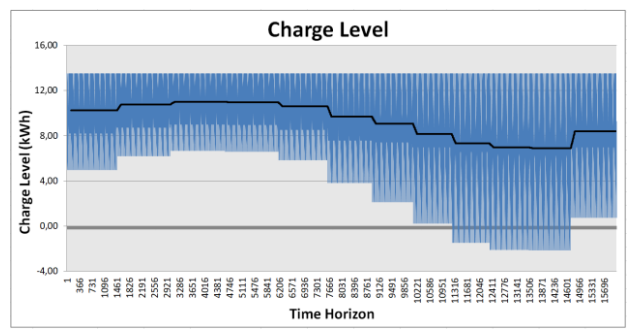

**Figure 15 - Raw Charge/Discharge profile**

The charge level must have an upper bound, but it cannot also be lower than zero. For this reason, the upper bound and thus the capacity must be adjusted to make sure that the lowest charge level is zero. The DOD of the Tesla batteries is 100% and thus this choice is justified. Using the Solver Parameters of Excel, the upper bound is changed to set the minimum value of the charge level equal to zero. The new value is 15,62 kWh which is more than the PowerWall 2 itself. The final profile is the same but moved up to match the lower bound with the value zero.

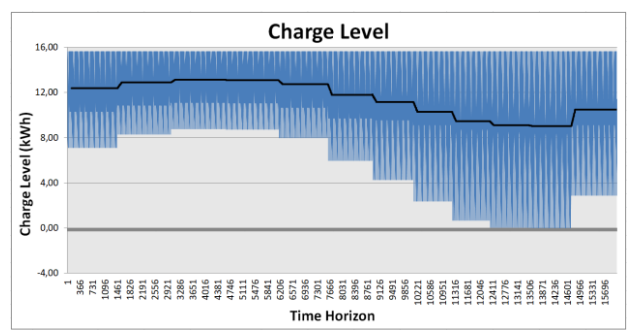

**Figure 16 - Corrected Charge/Discharge Profile**

The battery results in a capacity of 15,6 kWh and the overall mass is 1723 kg.

### 7.3 Non-Nuclear Configuration

In this first case, the use of all possible generators except the kilopower is accounted for.

This condition involves three different sub-configurations active in the sun season, storm season, and in the 7th month of Sagittarius, in which photovoltaic and turbine cannot have an effective production.

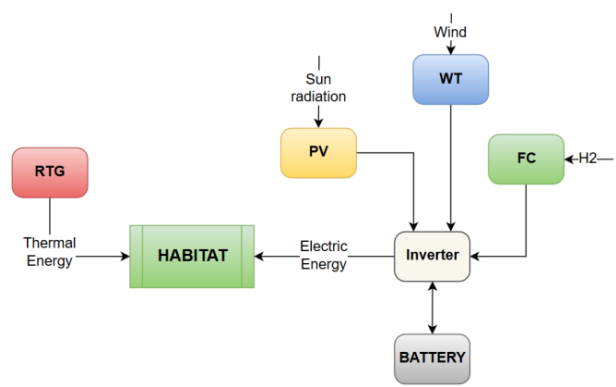

**Figure 17 – Non-Nuclear Configuration**

During the sun season, the turbine cannot be effective but the photovoltaics can be exploited at the best of their capacity. In this sense, the optimal condition is a surface of 330,5 m² of photovoltaics coupled with a 102,4 kWh battery. This configuration was "helped" using the fuel cell with constant production of 2,246 kWh/h, allocating 1000 kg of hydrogen over 1500 kg available from the ERV (6).

In the seventh month, the fuel cell is the main generator, taking into account that the photovoltaic is still able to produce between 1 and 3 kWe, despite the low solar radiation. The amount of hydrogen used in this period is quantified in 484,6 kg out of 500 kg left from the previous season.

Finally, in storm season an airborne turbine has been sized in relation to the battery, to cover the load and make sure that the minimum value of the battery is not less than zero. Be noted again that also in this case the photovoltaics can produce a useful amount of energy as well as the 7th month. For this reason, accounting for the photovoltaics is still important not to oversize the wind turbine, which results in a 3,52 m rotor.

The total mass of this energy system is 2975 kg, 925,4 of which are photovoltaic, 790,8 battery, 36,8 inverter, 45,9 fuel cell, 1028,2 turbine and 147,9 for the RTG which will entirely cover the thermal demand. The sizing of the RTG, in particular, quite straightforward since it is only necessary to consider the highest value in the thermal demand. Thermal storage is a viable way to undersize the RTG, but it was not considered in this case (also because changing a smaller RTG with storage would probably not decrease the mass of that particular part of the energy system).

The battery profile has been obtained in the same way as the full nuclear case and it will be shown here, for the sake of brevity.

## 8. Conclusions

It is clear that the Full Nuclear configuration is cheaper from different points of view:

- It saves 1252 kg compared to the non-nuclear configuration, with savings in the order of tens of million dollars, considering the average cost of sending one kilogram of material only to Low Earth Orbit, which is about 20.000$/kg (30).

- It allows satisfying both the thermal and electric needs without the need of several sources that compensate each other along the year.

- It does not depend on the environmental conditions, which are still difficult to predict, especially in the case of sandstorms, which can occur at any time during the storm season and last for months.

- Uranium (usually U238 enriched at the 3,5% of U235) has an energy density of several orders of magnitude higher than the conventional fuels so that few kilograms would be needed to run a reactor for years.

## Acknowledgements

This paper originates from the Author's Master Thesis and it has been developed originally in 2017-2018 in a joint collaboration between the University of Cagliari and the Waseda University of Tokyo. My thanks go to both supervisors who followed me in this work: Professor Andrea Frattolillo and Professor Yoshiharu Amano.

The following paper has been completed in the April 2021 with a few corrections and updates.

## References

1. ZUBRIN, Robert, BAKER, DAVID and GWYNNE, Owen. *Mars direct-A simple, robust, and cost effective architecture for the Space Exploration Initiative.* 1991. p. 329. 29th Aerospace Sciences Meeting.

2. Keener, Molly S. AndersonMichael K. EwertJohn F. *Baseline Values and Assumptions Document.* NASA Johnson Space Center. 2018. https://ntrs.nasa.gov/api/citations/20180001338/downloads/20180001338.pdf.

3. MILLOUR, Ehouarn, et al. *The Mars climate database (MCD version 5.2).* 2015. pp. 2015-2438. European Planetary Science Congress.

4. JAMES, George, CHAMITOFF, Gregory Errol and BARKER, Donald. *Resource utilization and site selection for a self-sufficient martian outpost.* National Aeronautics and Space Administration, Lyndon B. Johnson Space Center. 1998. NASA/TM-98-206538.

5. WANG, Huiqun and RICHARDSON, Mark I. *The origin, evolution, and trajectory of large dust storms on Mars during Mars years 24–30 (1999–2011).* s.l. : Icarus, 2015. pp. 112-127.

6. ZUBRIN, Robert and Richard, WAGNER. *The Case for mars.* s.l. : Simon and Schuster, 2011. ISBN: 978-0-684-82757-5.

7. Kennedy, Raboin, Spexarth, Valle. *Inflatable Structures Technology Handbook. Chapter 21: Inflatable Habitats.* NASA Johnson Space Center . 2000. http://citeseerx.ist.psu.edu/viewdoc/download?doi=10.1.1.730.8393&rep=rep1&type=pdf.

8. John R. Howell, M. Pinar Mengüç. *Radiative transfer configuration factor catalog: A listing of relations for common geometries.* s.l. : McGraw-Hill, 2011. pp. 910-912. Journal of Quantitative Spectroscopy and Radiative Transfer, Volume 112, Issue 5,.

9. GENDRON, Stephane, et al. *Phoenix mars lander mission: Thermal and cfd modeling of the meteorological instrument based on flight data.* Barcelona, Spain : s.n., 2010. p. 6195. 40th International Conference on Environmental Systems.

10. Holman, J.P. *Heat Transfer 10th Edition.* s.l. : McGraw-Hill, 2010. ISBN-10: 0073529362 .

11. CHATTERJEE, Sharmista. *Systems analysis of a closed loop ECLSS using the ASPEN simulation tool. Thermodynamic efficiency analysis of ECLSS components.* 1993. Document ID: 19940007894. https://ntrs.nasa.gov/citations/19940007894.

12. Met - Metabolic Rate. [online]. *Engineering ToolBox.* [Online] 2004. [Accessed 26/01/2018]. https://www.engineeringtoolbox.com/met-metabolic-rate-d_733.html.

13. STEIN, T. P., et al. *Energy expenditure and balance during spaceflight on the space shuttle.* s.l. : American Journal of Physiology-Regulatory, Integrative and Comparative Physiology, 1999. 276.6: R1739-R1748.

14. DREYER, Georges. *THE NORMAL BASAL METABOLISM IN MAN: AND ITS RELATION TO THE SIZE OF THE BODY AND AGE, EXPRESSED IN SIMPLE FORMULÆ .* s.l. : The Lancet, 1920. pp. 289-291. Volume 196, Issue 5058, ISSN 0140-6736.

15. MUSCATELLO, Anthony C., et al. *Testing and Modeling of the Mars Atmospheric Processing Module.* 2017. p. 5149. AIAA SPACE and Astronautics Forum and Exposition: Session: In-Situ Resource Utilization (ISRU).

16. V. Barbarossa, G. Vanga, G. Battipaglia. *Studio e sperimentazione di processi chimico-fisici di trattamento e conversione del syngas.* Agenzia Nazionale per le Nuove Tecnologie, Energia e lo Sviluppo Economico Sostenibile (ENEA). 2011. https://www.enea.it/it/Ricerca_sviluppo/documenti/ricerca-di-sistema-elettrico/carbone-pulito-e-ccs/RdS322.pdf/view.

17. SCHILD, P. G. and MYSEN, M. *Recommendations on specific fan power and fan system efficiency.* Air Infiltration and Ventilation Centre/INTERNATIONAL ENERGY AGENCY. 2009. Technical Note AIVC 65. https://www.aivc.org/sites/default/files/members_area/medias/pdf/Technotes/TN65_Specific%20Fan%20Power.pdf.

18. Brodrick, James. *Solid-State Lighting: R&D Plan.* U.S. Department of Energy. 2016. https://www.energy.gov/sites/default/files/2018/09/f56/ssl_rd-

plan_jun2016.pdf.

19. Hanna Dodd, Scott Farley, John Onderdonk. *Caltech Energy Assessment for Laboratories (CEAL).* California Institute of Technology. 2011. https://studylib.net/doc/18516547/progress-report---sustainability-at-caltech.

20. LICHTER, Matthew D. and VITERNA, Larry A. *Performance and Feasibility Analysis of a Wind Turbine Power System for Use on Mars.* National Aeronautics and Space Administration, Glenn Research Center. 1999. NASA/TMm1999-209390.

21. DELGADO-BONAL, Alfonso and MARTÍN-TORRES, F. Javier. *Solar cell temperature on Mars.* 2015. pp. 74-79. Solar Energy, Volume 118.

22. GIBSON, Marc A., et al. *NASA's Kilopower reactor development and the path to higher power missions.* 2017. pp. 1-14. IEEE Aerospace Conference 2017. .

23. BENNETT, Gary, et al. *Mission of daring: the general-purpose heat source radioisotope thermoelectric generator .* 2006. p. 4096. 4th International Energy Conversion Engineering Conference and Exhibit (IECEC).

24. Sigmaaldrich. Proton Exchange Membrane (PEM) Fuel Cells. [Online] Last Update 2021. https://www.sigmaaldrich.com/materials-science/renewable-alternative-energy/pem-fuel-cells.html.

25. PENEV, Michael. *Hybrid hydrogen energy storage.* NREL, All-Energy, Aberdeen, UK, U.S. Department of Energy. 2013.

26. Tesla. *Tesla Powerwall2 Brochure.* 2019. https://www.tesla.com/sites/default/files/pdfs/powerwall/Powerwall%202_AC_Datasheet_en_AU.pdf.

27. —. Powerpack. [Online] Tesla, Last updated 2021. https://www.tesla.com/it_IT/powerpack?redirect=no.

28. GIBSON, Marc A., et al. *The Kilopower reactor using Stirling TechnologY (KRUSTY) nuclear ground test results and lessons learned.* 2018. p. 4973. International Energy Conversion Engineering Conference. 2018.

29. Rucker, Michelle A. *Integrated Surface Power Strategy for Mars.* NASA, Lyndon B. Johnson Space Center. 2015. pp. 23-26. Nuclear and Emerging Technologies for Space.

30. Jones, Harry W. *The Recent Large Reduction in Space Launch Cost.* 2018. 48th International Conference on Environmental Systems.

31. LICHTER, Matthew D. and VITERNA, Larry A. *Performance and Feasibility Analysis of a Wind Turbine Power System for Use on Mars.* National Aeronautics and Space Administration, Glenn Research Center. 1999. NASA/TMm1999-209390.